\documentclass{aastex}
\usepackage{emulateapj5,danonecolfloat}

\shorttitle{CO Above the Disk of M82}
\shortauthors{Taylor, Walter \& Yun}

\def\kms{km\thinspace s$^{-1}$}
\def\cm2{cm$^{-2}$}
\def\deg{^{\circ}}

\begin{document}

\twocolumn[    % Begin \twocolumn for the emulateapj5 file

\title{Discovery of Molecular Gas in the Outflow and Tidal Arms around M82}

\author{Christopher L. Taylor}
\affil{Five College Radio Astronomy Observatory, University of
Massachusetts, 619 Lederle GRT, Amherst, MA 01003}
\author{Fabian Walter}
\affil{California Institute of Technology, Astronomy Department 105-24,
Pasadena, CA, 91125}
\and
\author{Min S. Yun}
\affil{Department of Astronomy, University of Massachusetts, 619 
Lederle GRT, Amherst, MA 01003}

%\offprints{C.L. Taylor}
%\date{Received / Accepted }
%\maketitle

\begin{abstract}

We present the first fully sampled map of $^{12}$CO (1-0) emission from 
M82 covering the entire galaxy.  Our map contains a $\sim$ 
12~$\times$15 kpc$^2$ area.  We find that extraplanar CO emission,  previously 
reported at short distances above the galactic plane, extends to heights of 
up to 6 kpc above the disk. Some of this emission is associated 
with tidal arms seen in HI, implying either that M82 contained substantial 
amounts of molecular gas in the outer disk, or that molecular gas formed {\it 
after} the tidal features.  CO emission along the direction of the outflow 
extends to distances of $\sim$ 3 kpc above and below the disk.  At this 
distance, the line is shifted in velocity $\sim 100$ \kms, and has the same 
sense as the galactic outflow from the central starburst.  This implies 
that molecular gas may be entrained into the outflow.

\end{abstract}


] %end front material encased in \twocolum

\section{Introduction}

M82 is one of the closest starburst galaxies (D = 3.63 Mpc assuming the 
distance of M81; \citet{Fe94}), and one of the most intensively studied; it 
has been observed at nearly every wavelength possible (\citet{LHW} summarize 
the past observations).  It is a well known prototype of the starburst--driven 
galactic superwind phenomenon \citep{AT}.  The galactic winds, created by 
the combined effect of supernovae and stellar winds \citep{TTB}, are thought 
to drive material up out of the disk.  M82 is also a member of a triplet 
system (with M81 and NGC~3077) undergoing a strong gravitational interaction 
\citep{YHL}.  Material can be drawn out of the disk by such interactions, 
creating tidal features up to tens of kiloparsecs in length (e.g. \citet{SSP}, 
\citet{I94}).  M82 features both a superwind and tidal arms.

M82 has been observed in various CO lines, both with single dish 
telescopes and interferometers (e.g. \citet{OR84}, \citet{YS84}, \citet{Ne87}, 
\citet{Se92}, \citet{SL}, \citet{We99}, \citet{We01}, and \citet{SC}), 
mostly focusing on the inner 1 kpc of the galaxy. In this paper we present 
the first fully sampled, wide--field map of the $^{12}$CO J = 1$\rightarrow$0 
emission in M82, covering an area of $\sim12\times15$ kpc (improving previous 
areal coverage by more than an order of magnitude) to search for molecular 
gas associated with either the superwind or the tidal features.

\section{Observations and Data Reduction}

The observations were obtained with the Five College Radio Astronomy 
Observatory 14~m telescope over several runs from January -- May, 2001. 
We observed the $^{12}$CO J = 1$\rightarrow$0 line using SEQUOIA, a focal 
plane array receiver, consisting of 16 pixels.  Because the beamsize 
of the telescope at 115 GHz is 44\arcsec\ and the spacing between pixels in 
the array is 88\arcsec, the beams were shifted in half-beam steps to
fill in the gaps and make a fully sampled map.  The extragalactic 
filterbanks were used to obtain a bandwidth of 320 MHz with 5 MHz (13 \kms) 
channels.  System temperatures ranged from $\sim$ 450 to 800 K.  The pointing 
was checked every 2 to 4 hours each session using SiO maser sources.

Six SEQUOIA footprints (6\arcmin~$\times$~6\arcmin\thinspace) were observed, 
partly overlapping, for a total area of 12\arcmin~$\times$~15\arcmin\ 
(Fig.~2). The data were reduced with the CLASS package.  We subtracted linear 
baselines from each spectrum and when a good fit didn't result, the spectrum 
was discarded.  Over 9,000 spectra were collected and coadded; the total 
integration time on source was $\sim$ 45 hours. A main beam efficiency of 
$\eta_{MB}$\,=\,0.45 was used used to put the spectra on the T$_{MB}$ scale.  
We smoothed the data spatially to 90\arcsec\ and in velocity to 39 \kms\ to 
emphasize low S/N extended structure.  Unless otherwise stated, all discussion 
refers to the smoothed data cube.

% Below is the figure information for the emulateapj5

\begin{figure}[tb]
\centerline{\epsfxsize=9.3cm\epsffile{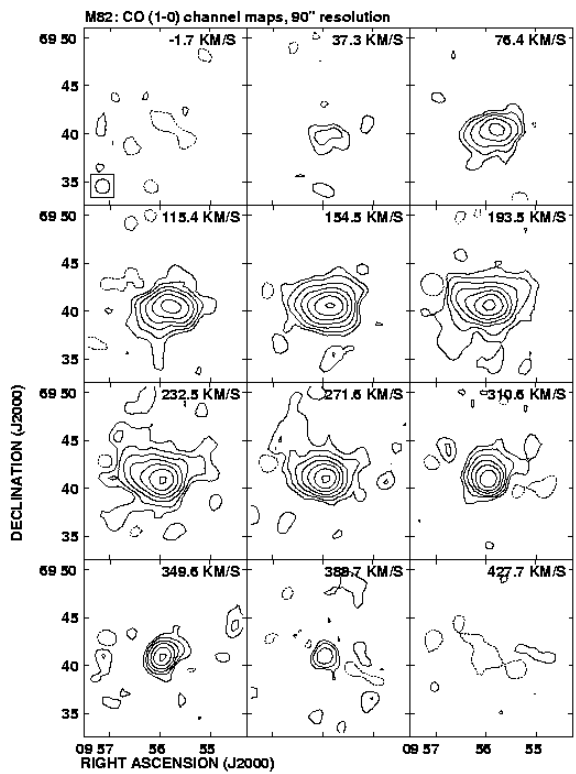}}
\caption{\footnotesize Channel maps showing $^{12}$CO (1-0) emission,in M82
smoothed to a resolution of 90\arcsec.   The contours are in units of -2,
2, 4, 8, 16, 32, 64 and 128$\sigma$ where $\sigma$ is the rms noise in a 
channel map, equal to 6.6 mK.  
\label{fig1} }
\vskip-0.5cm
\end{figure}

\section{Analysis and Results}

\subsection{The Global Distribution of the Gas}

Our observations reveal molecular gas at large heights above and below the 
plane of M82: Fig.~1 shows the channel maps,  and Fig.~2 the total integrated 
CO emission.  Given the strong emission from the center of M82 (0.76 K in 
our 60\arcsec, 39 \kms\ data), it is possible that any weak emission might 
be the response of the telescope error beam to the strong source.  A careful
inspection of our data shows this isn't a problem in our map for two
reasons: 1) the velocity of the extended emission is shifted relative to 
the center of M82 (Fig.~6), and 2) the response of the error beam would 
be 100 to 1000 times weaker than the peak emission, but the peak of the 
extended emission is only about 15 times weaker.

North of the galaxy, molecular gas is found up to 6 kpc (all values are 
deconvolved for the 90\arcsec\ beam, 1\arcmin\ $\sim$ 1 kpc for the adopted 
distance of 3.63~Mpc) away from the plane of the disk.  More emission is 
detected $\sim$4\ kpc above the disk at $\sim$ 2 kpc west of the minor axis. 
Along the minor axis, emission reaches heights of $\pm\sim$ 3 kpc. The two 
features seen towards the south are probably not real (see the discussion 
below).

The total CO integrated intensity is 6.1~$\times~10^2$ K \kms. Assuming a 
Galactic CO--to--H$_2$ conversion factor of $2.3~\times~10^{20}$ cm$^{-2}$ 
(K \kms)$^{-1}$ \citep{Se88} yields $4.3~\times~10^{9}$ M$_{\odot}$ (but see 
\citep{We01} for a lower conversion factor in M82).  This compares to 
the value of $1.7~\times~10^{9}$ M$_{\odot}$ obtained by \citep{YS84} (after 
correcting to our assumed distance and conversion factor).  The difference 
between our value and theirs results from several factors: 1) they did not 
convert to T$_{mb}$ so with an $\eta_{MB}$ of 0.5 at that time, the two 
values agree to within 20\%; 2) they didn't map the entire galaxy; and 3) 
they had a sparse sampling away from the galactic disk.  

To determine how much emission originates outside the disk of M82,
we measured the FWHM of the emission along the minor axis ($\sim$ 120\arcsec) 
and defined the extraplanar emission to be that which originates at a 
distance of at least 120\arcsec\ from the major axis.  This flux adds up to 
$\sim$ 10\% of the total.  Because our large beam doesn't resolve the 
disk emission, it is hard to decompose this properly, so this number should 
be taken with caution.

% Below is the figure information for the emulateapj5

\begin{figure}[tb]
\centerline{\epsfxsize=9.3cm\epsffile{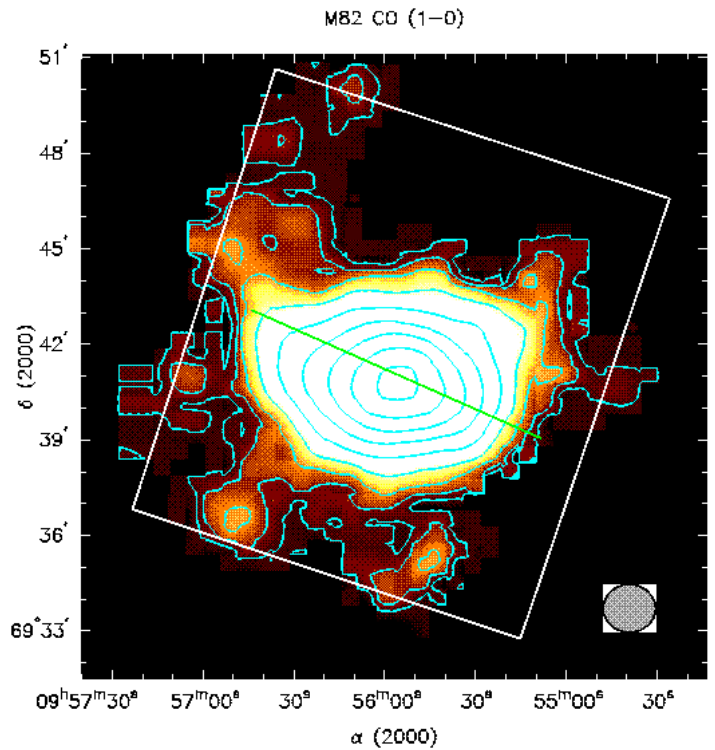}}
\caption{\footnotesize An integrated intensity image and contours of
$^{12}$CO (1-0) emission in M82, smoothed to 90\arcsec\ resolution. The   
 data cube was blanked at the 2$\sigma$ level prior to making this image. 
 The contours are 0.31, 0.63, 1.25, 2.5, 5, 10, 
 20, 40 and 80\% of the peak value (1.0$~\times~10^5$ K \kms beam$^{-1}$.  
 The green line shows the plane of the galaxy disk.  The beam size is 
 indicated in the lower right.  The box shows the region mapped, where
 emission contours extend beyond the box because we regridded and spatially
 smoothed the data.
\label{fig2} }
\vskip-0.5cm
\end{figure}

\subsection{Kinematics}

Some of the CO emission is clearly associated with tidal features.  Fig.~3 
shows the CO (1-0) contours superposed over an HI map.  The HI map combines 
unpublished B, C \& D configuration data from the NRAO-VLA\footnotemark 
\footnotetext{ The National Radio Astronomy Observatory is a facility of 
the National Science Foundation operated under cooperative agreement by 
Associated Universities, Inc} with a 21\arcsec\ $\times$ 19\arcsec beam.  
The HI arm projecting off the northeast (NE) end is the most obvious tidal 
feature, and the molecular gas here is associated with this arm.  To 
the northwest (NW) of the galaxy, the CO emission closely follows a set of
HI clouds. HI and CO spectra for both regions (NE and NW) are
presented in Fig. 4. The HI is not situated along the axis of outflow,
so both the CO and HI emission are probably of tidal origin. The NW feature 
lies closer to the minor axis of M82 and may be influenced by the starburst 
driven outflow.  In the south (S1) the CO contours seem to follow 
the HI closely in the spur projecting more than 300\arcsec\ southeast of 
the galaxy center.  However, as Fig.~4 shows, the velocities of the HI 
and CO don't match here. The cloud to the very south (S2) has
no corresponding HI emission at any velocity.  These two southern features 
(and the eastern edge of NE) fall within one 90\arcsec\ beam of our
map edge.  Because there is no data beyond the map edge, smoothing 
spatially will increase the noise along the edge, relative to the 
interior of the map; these edge features may be artifacts.

% Below is the figure information for the emulateapj5

\begin{figure}[tb]
\centerline{\epsfxsize=9.3cm\epsffile{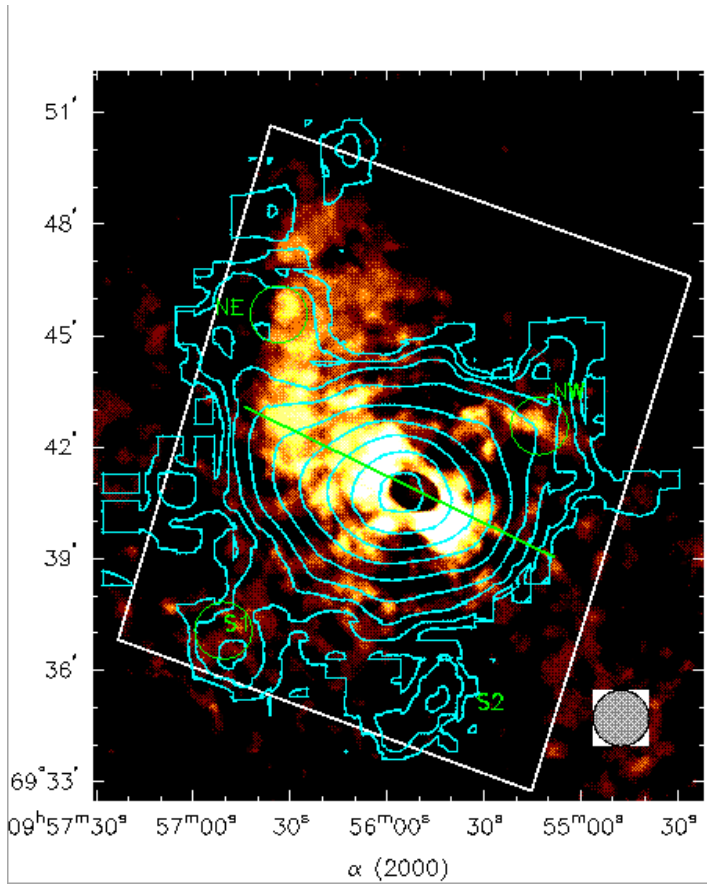}}
\caption{\footnotesize $^{12}$CO (1-0) contours superposed on a VLA HI
integrated intensity map.  The CO contours are the same as in Fig.~2,
and the CO beam size is shown in the lower right.  The white box indicates
the region mapped in CO.  The green line indicates
the major axis of M82.  The circles show the areas where the CO and HI
spectra are compared in Fig.~4.  
\label{fig3} }
\vskip-0.5cm
\end{figure}

% Below is the figure information for the emulateapj5

\begin{figure}[tb]
\centerline{\epsfxsize=9.3cm\epsffile{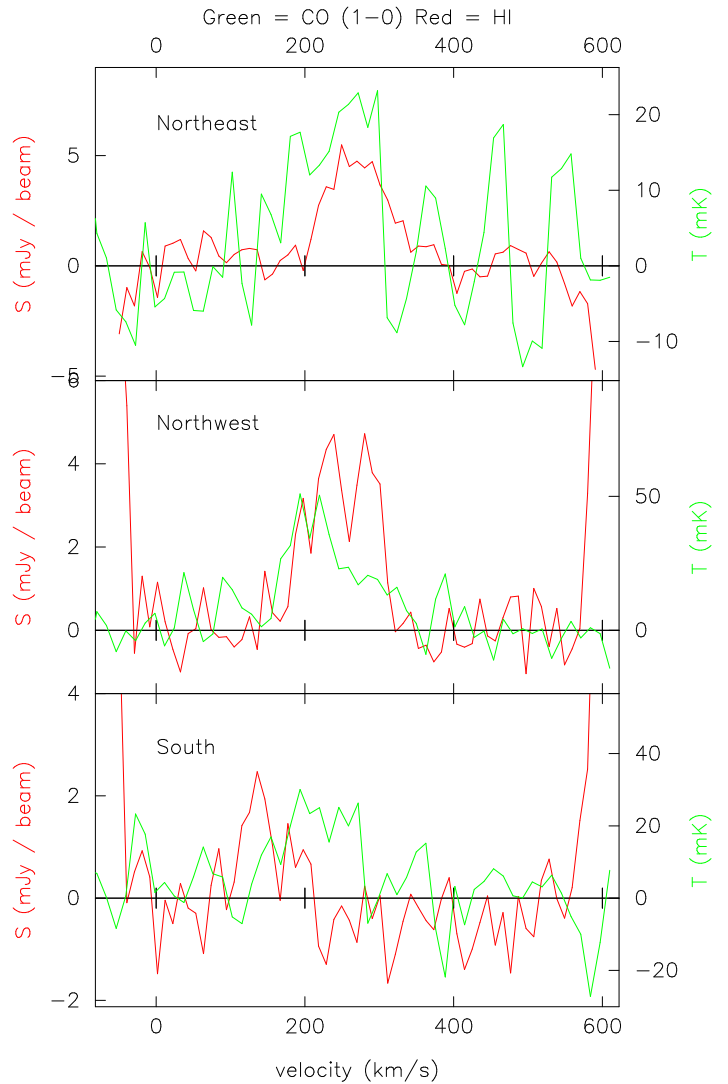}}
\caption{\footnotesize Spectra through the CO and HI data cubes for the
three tidal features in the text.  The HI is plotted in red, the CO in
green.
\label{fig4} }
\vskip-0.5cm
\end{figure}

While some of the CO emission away from the disk is of tidal origin,
some participates in the well known outflow in M82.  Previous 
observations found that CO near the major axis is outflowing ($\pm$ 
42\arcsec\ in \citet{SC}; $\pm$ 45\arcsec\ in \citet{Ne87} ), but 
we show for the first time that the molecular outflow extends to $\sim$ 3 
kpc (180\arcsec).  Fig.~5 shows the CO contours superposed on a 
continuum subtracted H$\alpha$ image obtained from the Astronomical Digital
Image Library (http:$\slash\slash$adil.ncsa.uiuc.edu$\slash$).  Excluding 
the tidal features, the shape of the CO contours is similar to the outflow 
traced in H$\alpha$. However, due to our low spatial resolution we cannot 
correlate the CO emission to specific H$\alpha$--filaments.  In any event, 
it is not clear that a correlation is expected as the molecules would
be dissociated before the gas is ionized.

% Below is the figure information for the emulateapj5

\begin{figure}[tb]
\centerline{\epsfxsize=9.3cm\epsffile{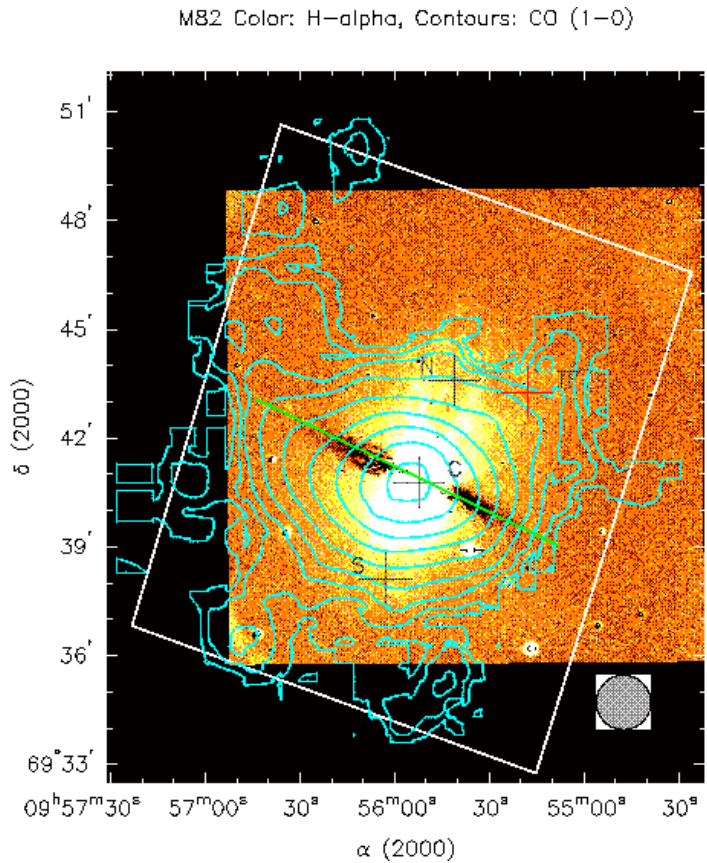}}
\caption{\footnotesize $^{12}$CO (1-0) contours superposed on a continuum
subtracted H$\alpha$ map.  The CO contours are from Fig.~2.
The CO beam size is shown in the lower right.  The green line indicates
the major axis of M82. The crosses show the areas compared in Fig.~6
\label{fig5} }
\vskip-0.5cm
\end{figure}

Fig.~6 shows evidence that molecular gas is affected by the galactic 
outflow.  We show three spectra from along the minor axis of M82 -- north (N)
and south (S) of center, and at the center (C) (indicated in Fig.~5).  The 
offset in velocity between the S and C spectra 
is obvious and amounts to a difference of 100 \kms.  The N spectrum is very 
wide, however, and lacks a central peak, making it hard to discern an offset 
compared to the galaxy center.  It appears to contain not only emission from 
the outflow (at $\sim$ 300 \kms), but contamination from the northwest tidal 
feature (TF).  We plot in red the TF spectrum over the emission from N,
showing that the emission at lower velocities at N is similar in velocity 
range to that from TF.  Obtaining an outflow velocity here is problematic; 
due to the contamination from TF it is impossible to isolate the minor axis 
emission.  If we assume the minor 
axis outflow contribution dominates at the extreme redshifted range of 
velocities, then we obtain a value of $\sim$ 80 \kms, but this should be 
taken with a grain of salt.  The redshift of S and possible blueshift
of N have the same sense as observed in other lines (e.g.  CO (3--2) by 
\citet{SC}; H$\alpha$ by \citet{McKe95, SB}).   This is generally attributed
to a large scale outflow with an inclination of 80$\deg$ to the disk. 

% Below is the figure information for the emulateapj5

\begin{figure}[tb]
\centerline{\epsfxsize=9.3cm\epsffile{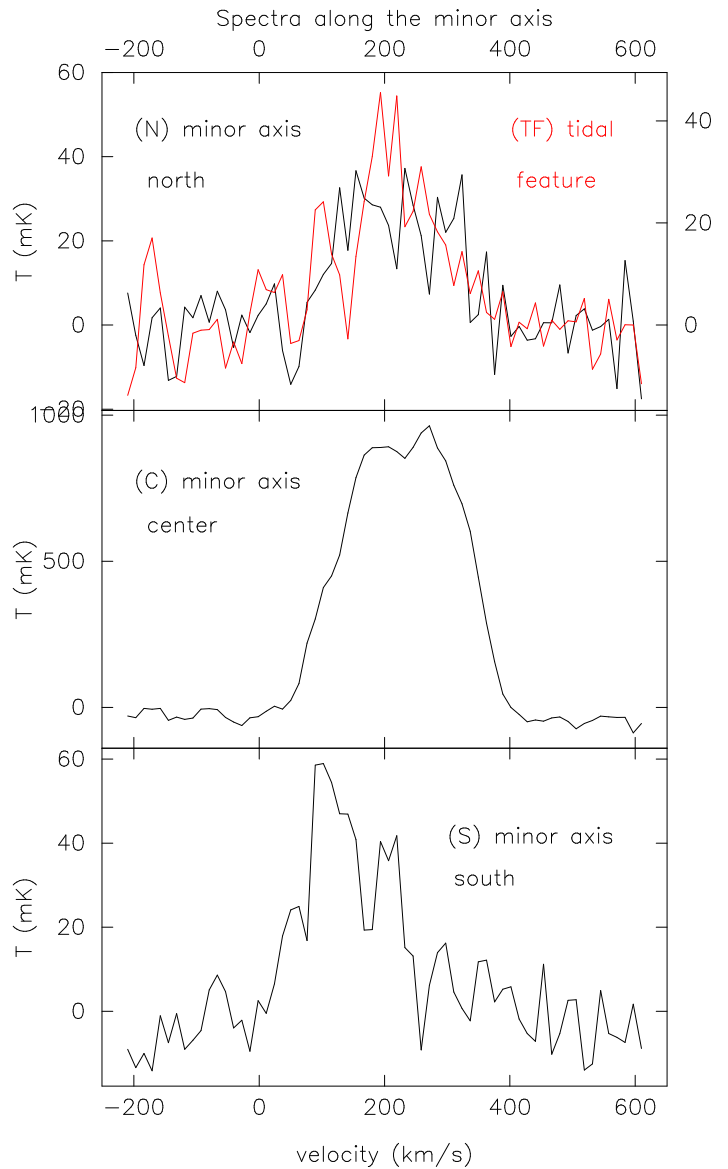}}
\caption{\footnotesize 
Spectra taken from positions along the minor axis.
{\it Bottom:} the southern end of the minor axis.  {\it Center:} the center
of the galaxy.  {\it Top:} the northern end of the minor axis, with a
spectrum from the northwest tidal feature plotted in red for comparison.
\label{fig6} }
\vskip-0.5cm
\end{figure}

\section{Summary}

We present the first complete, fully sampled map of the $^{12}$CO (1-0) 
emission in the prototypical starburst galaxy M82, covering an area of $\sim$ 
12~$\times$~15 kpc$^2$.  M82 is nearly edge on ($i = 80\deg$) allowing us 
to search for emission above the disk; we detect CO to great heights out of 
the galactic plane (up to 6 kpc).  Some of this extraplanar molecular gas 
is in tidal arms generated by the interaction with M81.  Gas in tidal arms 
normally comes from the outer disk regions, so that we see CO emission in 
the arms means that M82 either had molecular gas in its outer disk, or else 
the molecular gas formed {\it in} the tidal arms after the interaction.  
Either way, it also requires sufficient metal abundance to form the observed 
CO in addition to the implied H$_2$.
We see emission along the minor axis $\sim$ 3 kpc above and below the disk.  
In the north {\it some} of this emission may be associated with tidal features,
but in the south it is not.  The velocity difference of the CO emission along
the minor axis has the same direction as what is seen in the H$\alpha$ 
emission line.  This CO emission likely comes from molecular gas which has 
been entrained in the starburst driven outflow in M82.

\begin{acknowledgements}
We thank the referee for helpful comments that improved this paper.
The Five College Radio Astronomy Observatory is operated with the permission 
of the Metropolitan District Commission, Commonwealth of Massachusetts, and 
with the support of the National Science Foundation under grant AST-9725951.
FW acknowledges NSF grant AST 96-13717.

\end{acknowledgements}

\end{document}